\documentclass[prl,twocolumn,superscriptaddress,showpacs]{revtex4}
\usepackage{bm}
\usepackage{graphicx,epsfig}
\usepackage{pstricks,multido,pst-3d,pst-node}
\usepackage{array}

\begin{document}

\title{Nanometer-resolved collective micromeniscus oscillations
  through optical diffraction}

\author{Helmut Rathgen}
\email{helmut.rathgen@web.de}
\affiliation{Physics of Complex Fluids, University of Twente, The Netherlands.}
\affiliation{J.M.\ Burgers Centre of
  Fluid Dynamics and MESA$^+$- and IMPACT-Institutes, University of Twente, The Netherlands.}
\author{Kazuyasu Sugiyama}
\affiliation{Physics of Fluids, University of Twente, The Netherlands.}
\affiliation{J.M.\ Burgers Centre of
  Fluid Dynamics and MESA$^+$- and IMPACT-Institutes, University of Twente, The Netherlands.}
\author{Claus-Dieter Ohl}
\affiliation{Physics of Fluids, University of Twente, The Netherlands.}
\affiliation{J.M.\ Burgers Centre of
  Fluid Dynamics and MESA$^+$- and IMPACT-Institutes, University of Twente, The Netherlands.}
\affiliation{School of Physical and Mathematical Sciences
, Nanyang Technological University, Singapore.}
\author{Detlef Lohse}
\affiliation{Physics of Fluids, University of Twente, The Netherlands.}
\affiliation{J.M.\ Burgers Centre of
  Fluid Dynamics and MESA$^+$- and IMPACT-Institutes, University of Twente, The Netherlands.}
\author{Frieder Mugele}
\affiliation{Physics of Complex Fluids, University of Twente, The Netherlands.}
\affiliation{J.M.\ Burgers Centre of
  Fluid Dynamics and MESA$^+$- and IMPACT-Institutes, University of Twente, The Netherlands.}

\begin{abstract}
We study the dynamics of periodic arrays of micrometer-sized
liquid-gas menisci formed at superhydrophobic surfaces immersed
into water. By measuring the intensity of optical diffraction peaks in
real time we are able to resolve nanometer scale oscillations of
the menisci with sub-microsecond time resolution. Upon driving the
system with an ultrasound field at variable frequency we observe a
pronounced resonance at a few hundred kHz, depending on the exact
geometry. Modeling the system using the unsteady Stokes
equation, we find that this low resonance frequency is
caused by a collective mode of the acoustically coupled oscillating
menisci.
\end{abstract}

\pacs{47.55.dd,43.35.+d,42.25.Fx}

\date{\today}

\maketitle

Superhydrophobic surfaces have attracted much theoretical and
experimental interest \cite{quere05rpp68_2495}.
The entrapment of gas at the textured surface reduces the
actual interfacial area between the solid and the liquid,
which is at the origin of the distinctive properties of these
materials, including the large contact angle, the low
contact angle hysteresis\cite{lafuma03nmat2_457},
the self-cleaning
effect\cite{nakajima00lang16_7044}, and the large hydrodynamic
slip\cite{joseph06prl97_156104}.
The superhydrophobic state is closely related to
the shape of the microscopic liquid-gas interfaces spanning between the
ridges of the texture \cite{journet05epl71_104}. In this letter we
characterize the {\em dynamics} of these micromenisci and
introduce an optical diffraction measurement that reveals their
nanoscale motion.
We measure in real time
the optical diffraction intensity from a periodic array of
micromenisci. Driving the system with an ultrasound field at variable
frequency, we measure its frequency
response and we identify a well-defined
resonance peak with a center frequency well below the expectations for a
single micromeniscus. Modeling the system using the unsteady Stokes
equation and monopole interaction, we show that this frequency
reduction is due to acoustic coupling between menisci.

\begin{figure}
\rput(-1mm,29mm){a)}
\rput(46mm,29mm){b)}
\scalebox{0.7}{\includegraphics{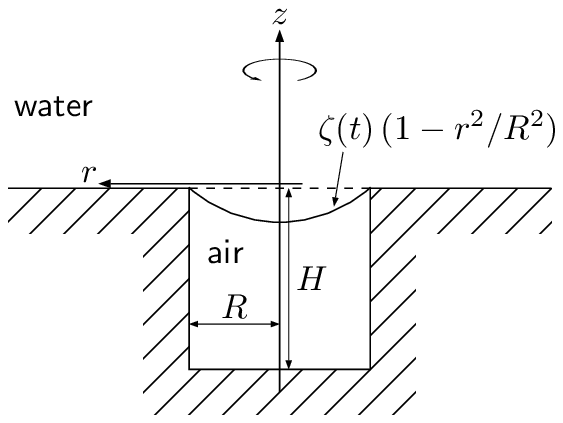}}
\hspace{4ex}
\scalebox{0.8}{\includegraphics{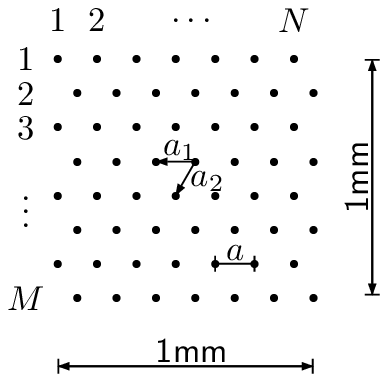}}
\caption{a) Schematic figure of a single micromeniscus. b)
  Pattern of an array of micromenisci. Arrows denote the primitive
  translations and nearest neighbor distance $a$.}
\label{fig:geometry}
\end{figure}

Fig.\ref{fig:geometry} shows the geometry of our system. It
consists of $1\times1$mm$^2$ wide hexagonal and square arrays of
micrometer-sized cylindrical holes (radii $R=2\mu$m or $3\mu$m,
depth $H=15\mu$m, nearest neighbor distance $a=15\mu$m or $25\mu$m).
The samples were fabricated from Si (110) using standard micro
lithography. Subsequently, the surfaces were hydrophobized by
vapor deposition of a monolayer of
1H,1H,2H,2H-Perfluorodecyltrichlorosilane,
following \cite{mayer00jvstb18_2433}. The advancing and receeding contact
angles on an unstructured surface are $\gamma_a=116^\circ$ and $\gamma_r=104^\circ$.
Upon immersing the samples into demineralized
water, ambient air is entrapped in every
hole, leaving a water-air meniscus behind that is pinned at the ridge of the
hole. Owing to the hydrostatic pressure, the menisci are bent inwards
with equilibrium curvature
$\kappa_0=\rho g h/\sigma$, where $g$ is the
acceleration of gravity, $\sigma$ is the water surface tension, and
$h\approx0.1$m is the distance between the sample and the free water
surface. This implies that the system is in diffusion
equilibrium and the gas pressure in the hole is the ambient pressure.
\begin{figure}
\raggedright
\rput[tl](0mm,60mm){a) Setup}
\scalebox{0.8}{\includegraphics{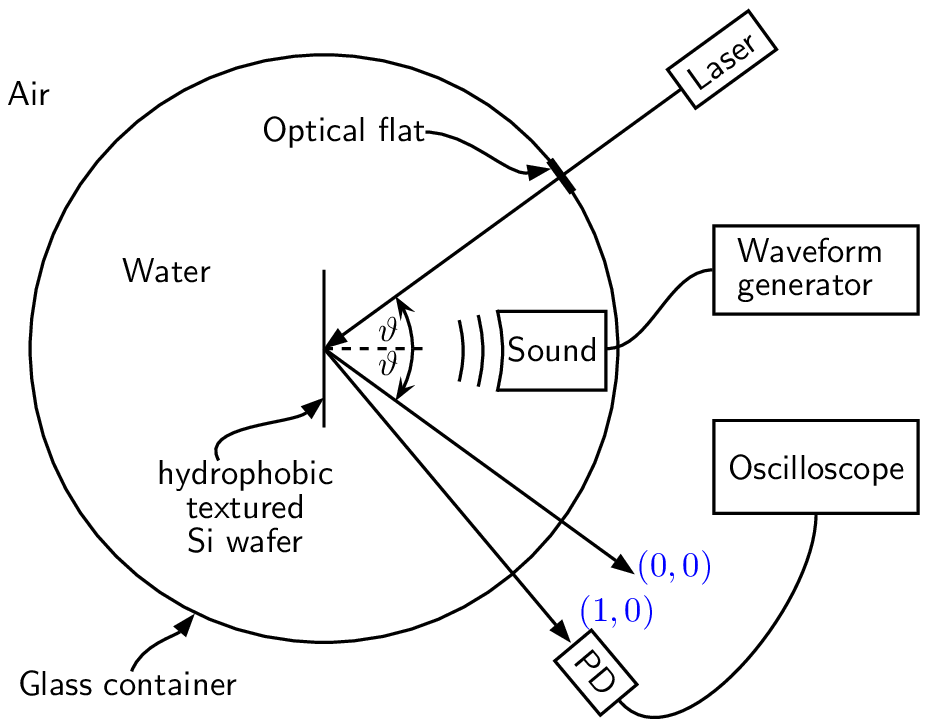}}\\[3ex]
\rput[tl](0mm,33mm){b) Diffraction pattern at $60^\circ$}
\scalebox{0.3}{\includegraphics[bb=72 436 456
      720]{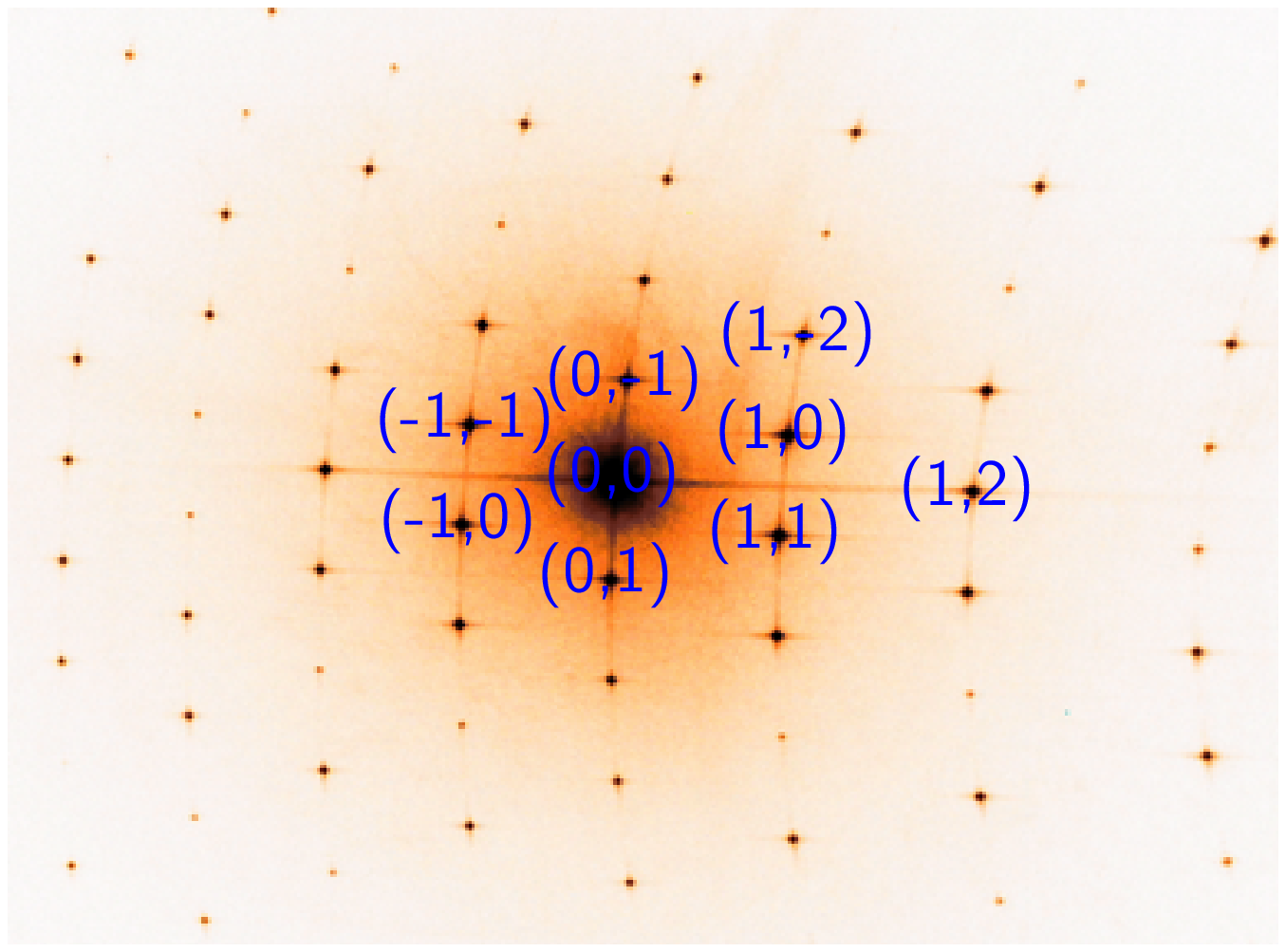}}
\hspace{2em}
\rput[tl](-3mm,33mm){c) Ultrasound response}
\includegraphics[scale=0.8,bb=60 52 168 162]{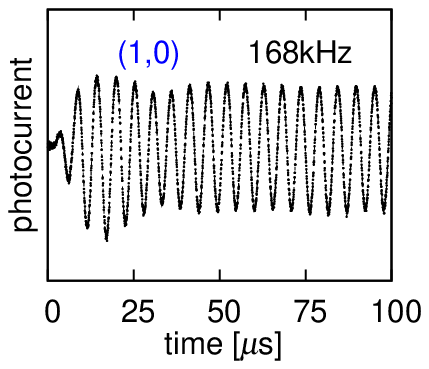}
\caption{a) Schematic figure of the experiment. b) Inverted
  photograph of the diffraction pattern ($\vartheta\approx
  60^\circ$). Numbers indicate Miller
  indices of diffraction orders. c) AC component of the
  light intensity measured
  in the (1,0) diffraction order at the beginning of an ultrasound
  wavetrain.}
\label{fig:setup}
\end{figure}
An Ar-ion laser ($\lambda=488$nm, s-polarized) is used to
illuminate the sample under an
angle typically between $60^\circ$ to $70^\circ$ with respect to
normal incidence (see Fig.\ref{fig:setup}).
The diffracted intensity is measured with a
photodiode positioned at a selected diffraction
peak, typically chosen in the vicinity of the specular reflected beam.
A broadband piezoelectric ultrasound transducer
is placed at its focal distance from the sample. The ultrasound
transducer is excited to emit finite wavetrains
using an arbitrary function generator. The ultrasound pressure at the
sample is of the order
$10^2$ to $10^3$ Pa, which is small compared to a critical static
pressure above which filling occurrs $P_c = 2\sigma\cos(\gamma_a)/R\approx
2.1\cdot10^4$Pa. To check the dynamic stability of the menisci, we
increased the ultrasound pressure to much larger values and observed
how the intensity oscillations disappeared at a defined threshold. The
ultrasound pressure is kept constant during a
frequency sweep by controlling the driving voltage according to the
transducers frequency response. 
Fig.\ref{fig:setup}c shows typical raw data corresponding to the
beginning of a wavetrain. After a transient lasting for a few
oscillation cycles, the signal becomes sinusoidal with a
constant amplitude. This amplitude is extracted from the
raw data by calculating the root mean square.

\begin{figure}
\rput(0mm,33mm){a)}
\includegraphics[scale=0.9]{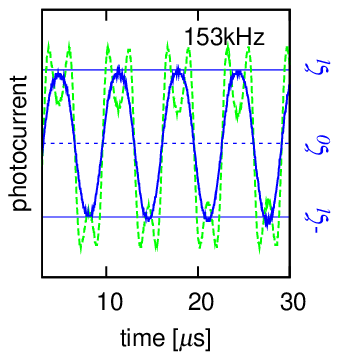}
\rput(1mm,33mm){b)}
\includegraphics[scale=1]{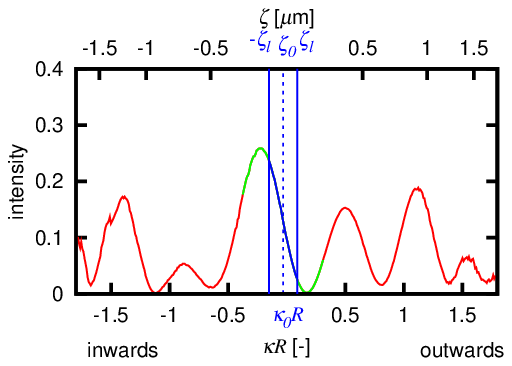}
\caption{a) Time-resolved intensity of the $(1,0)$
  diffraction order, corresponding to small ultrasound
  pressure ($\Delta P\approx 400$Pa), solid line, and large ultrasound
  pressure ($\Delta P\approx 800$Pa), dashed line.
  b) Calculated intensity of the first diffraction
  order as a function of meniscus deflection for a rectangular
  surface profile with groove width $w=2R$ ($\vartheta=66^\circ$). The
  data is displayed simultaneously in terms of the
  displacement $\zeta$ and 
  the non-dimensionalized curvature $\kappa R$. Both are related
  geometrically by
  $\zeta=2/\kappa(1-\sqrt{1-R^2\kappa^2/4})$. The equilibrium curvature
  $\kappa_0R=0.04$ corresponds to $\zeta_0=30$nm.}
\label{fig:workingpoint}
\end{figure}

The diffracted intensity depends in a highly nonlinear
way on the deflection of the menisci. To assure a linear
relation between the actual deflection and the measured intensity, the
menisci oscillations have to be small. This can be seen
in Fig.\ref{fig:workingpoint}a. While the diffracted intensity follows the
sinusoidal driving pressure at low driving amplitudes (solid line),
it is distorted at larger driving amplitudes (dashed line).
Prior to an experiment, we reduce the
ultrasound pressure until the undistorted sinus is observed.

To find the magnitude of the corresponding meniscus
deflections, we
consider the optical diffraction of the sample. In the
Fraunhofer limit, the diffracted
intensity is proportional to the intensity scattered by a single unit
cell, and for each unit cell, the
scattered intensity is governed by the interference of the elementary
waves emitted from the cell volume
\cite{born99principles_of_optics}. In our case the incident angle is
large and no light reaches the
bottom of the holes such that the interference takes place between the
waves emitted from the plane silicon
surface and the waves emitted from the meniscus. Qualitatively, as the
meniscus is
deflected -- consider the position $\zeta$ of the apex of the meniscus --
the intensity of a diffraction order (with diffraction angle $\approx$
incident angle) changes sinusoidally with a
period $T=\lambda/(2n\cos(\vartheta))$, where $n=1.33$ is the refractive
index of water.
To analyze these simple observations in detail, we performed a diffractive
optics calculation using the multilayer rigorous coupled wave analysis 
in the formulation of \cite{moharam95josaa12_1077}. This method allows
for calculating an exact solution to the Maxwell equations for the
optical response of arbitrary periodic surface profiles.
In Fig.\ref{fig:workingpoint}b we show the
resulting diffraction intensity as a function of the meniscus
deflection. The
typical distance between two
adjacent peaks corresponds to the period evaluated from the simple
Fraunhofer arguments above.
The result shows that the diffracted intensity is indeed linear in the
meniscus deflection in a range $[-\zeta_l..+\zeta_l]$ around the meniscus
equilibrium
position $\zeta_0$ and we find $\zeta_l\approx 90$nm.
Note that the extent of the linear range depends on the
incident angle $\vartheta$. It is larger for
larger incident angles, as can be seen readily from the simple
expression for $T$. Thus, the large angles that are used in
experiment allow for large
meniscus oscillations. For incident angles
above the angle of
total reflection between water and air $48.6^\circ$,
in addition the relative contribution of the menisci to the
scattered intensity is large.

The theoretical result for the linear
range is the key
to convert the measured intensity variations into absolute meniscus
deflections.
Since we have to assure linearity between intensity
and meniscus
deflection at {\em all} ultrasound frequencies, the
peak of the
resonance curve shown in the following has the height $\zeta_l$.
The slope of the linear range together with the
relative noise of the photodiode determines the resolution of the
deflection measurement. It is of the order 1nm under given
conditions.

\begin{figure}
\includegraphics[scale=0.7,bb=88 76 442 320,clip]{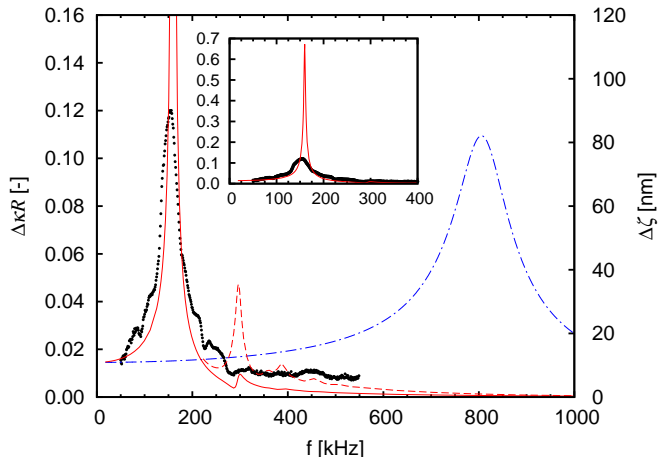}
\caption{Frequency response of an array of
micromenisci with hexagonal pattern, $a=15\mu$m and $R=3\mu$m. Crosses
show experimental data. The dash-dotted line
shows the theory for a single meniscus. Theoretical data for the
array is displayed in terms of
the mean (solid line) and the root mean square deflection (dashed
line).}
\label{fig:d6p15H}
\end{figure}

Fig.\ref{fig:d6p15H} shows a typical measured frequency response.
The sample displays a single resonance at
$f_r=153\pm5$kHz. Similar curves were obtained for all samples.
Table~\ref{tab:comparison} shows that the observed resonance
frequency increases both with increasing nearest neighbor distance and
with decreasing hole size.

\begin{table}
{\small
\setlength{\extrarowheight}{1pt}
\begin{tabular}{r|r|crr|r|r|r}
$R [\mu\textrm{m}]$&$a [\mu\textrm{m}]$&pattern&$N$&$M$&
$f_s$[kHz]&$f_c$[kHz]&$f_r$[kHz]\\
\hline
3 &15 &H&66&76&805&159&153$\pm$5\\
3 &15 &S&67&67&805&170&177$\pm$5\\
3 &25 &H&40&46&805&258&230$\pm$20\\
3 &25 &S&41&41&805&275&240$\pm$20\\
2 &15 &H&66&76&1433&346&285$\pm$20\\
2 &15 &S&67&67&1433&368&290$\pm$20\\
2 &25 &H&40&46&1433&549&400$\pm$20\\
2 &25 &S&41&41&1433&584&410$\pm$20\\
\end{tabular}
}
\caption{Resonance frequencies of the
  lowest collective mode $f_c$ as a
  function of lattice constant, pattern geometry and menisci
  radius. 'H' and 'S' denote hexagonal and square pattern,
  respectively. $f_s$ denotes the resonance frequency of a
  corresponding single meniscus, $f_r$ the experimental result.}
\label{tab:comparison}
\end{table}

To understand the observations quantitatively, we consider first the
response of a single meniscus under the influence of the
pressure field $P(t)=P_0+\Delta P e^{i2\pi f t}$, where
$f$ is
the ultrasound frequency, $P_0$ is the ambient pressure and $\Delta P$ is
the amplitude of the ultrasound pressure. Since
$\zeta_0\ll\Delta\zeta,R$, as follows with the values evaluated above, we
approximate the meniscus as flat
in equilibrium and its deflections as parabolic
$\zeta(r)=\zeta(1-r^2/R^2)$. The parabolic shape implies
that the curvature is uniform up to
$O(\zeta^3)$ giving right to the Laplace law. As
described in \cite{landau59fluid_mechanics}, the
smallness of the deflections $\Delta\zeta\ll R$, and the high frequency
$f\gg \nu/R^2$ of the oscillations ($\nu$ kinematic viscosity), allow
us to omit the non-linear term in the Navier-Stokes equation and
hence the dynamics of
the system is governed by the unsteady Stokes equation. The system
can then be described as a harmonic oscillator
\begin{equation}
\left\{\Phi(\omega^*)+{\rm
i}\Psi(\omega^*)+K\right\}\Delta\zeta^*=
-\Delta P^*. \label{eq:trans01}
\end{equation}
The transfer functions $\Phi$ and $\Psi$ account
for the inertia and the viscous damping due to the oscillatory flow
fields. All quantities are non-dimenionalized
$\omega^*=2\pi f R^2/\nu$, $\Delta\zeta^*=\zeta/R$ and $\Delta P^*=\Delta
P R^2/(\rho \nu)$. Since polytropic
and thermal dissipative effects \cite{chen98jasa104_1389} can be
neglected, the potential term reduces to a dimensionless spring constant
$K=R/(\rho\nu^2)\left(P_0R^2/(2H)+4\sigma\right)$. The first term with
the ambient pressure is due to the isothermal compression of the gas
and is negligble throughout this work, and the second
term with the surface tension $\sigma$ is
due to the surface energy of the liquid-gas interface.
The computation of $\Phi$ and $\Psi$ is performed by solving
the unsteady Stokes equation in cylindrical coordinates with classical
no-slip and free slip boundary conditions at the solid-liquid and
the liquid-gas interface, respectively. A detailed account of the
calculations will be given elsewhere \cite{rathgen06unpubjfm}.
While analytical expressions can be found both for the high
and for the low frequency limit, $\Phi$ and $\Psi$ have to be
computed numerically in the intermediate frequency range $10^0 <
\omega^* < 10^2$, which is relevant for the present experiments. The
dash-dotted line in Fig.\ref{fig:d6p15H} shows the solution for a
$R=3\mu$m hole with physical parameters of water
$\rho=10^3$kg/m$^3$, $\nu=10^{-6}$m$^2$/s,
$\sigma=73\cdot10^{-3}$N/m and ultrasound
pressure $\Delta P=390$Pa as used in the respective experiment. The
obtained resonance frequency is
approximately five times larger than the one observed experimentally.
Clearly, the single meniscus theory fails to describe the dynamics
of the system.

To resolve the
discrepancy, we consider the dynamic
coupling between the menisci. As the menisci oscillate in the external
pressure field, they emit pressure waves that affect
the other menisci. Denoting the non-dimensionalized distance between the
$i$-th and the $j$-th meniscus by $d_{ij}^*=d_{ij}/R$, the additional
force acting on the $i$-th meniscus can be expressed in terms of a
multipole expansion
$-\Delta\zeta_j^*{\omega^*}^2/(4d_{ij}^*)+O({d_{ij}^*}^{-2})$, where
${\Delta\zeta_j}^*$ is the deflection amplitude of the $j$-th
meniscus (see e.g., \cite{leroy05eupe17_189},
\cite{bremond06prl96_224501}).
To analyze the dynamics of the entire meniscus array, we extend the
equation of motion of the single meniscus, Eq.(\ref{eq:trans01}), by the
additional forces generated by all other menisci keeping only the
monopole term. We arrive at the coupled equations of motion
\begin{equation}
\left\{\Phi(\omega^*)+{\rm i}\Psi(\omega^*)+K\right\}\Delta\zeta_i^*=
-{\Delta P}^* +\sum_{i\neq j}^{N\cdot M}
\frac{\omega^{*2}}{4d_{ij}^*}{\Delta \zeta_j}^*. \label{eq:trans02}
\end{equation}
The coupling term gives rise to an additional effective mass, which
reduces the resonance frequency, as required.
We solve Eq.\ref{eq:trans02} for the individual deflection
amplitudes
${\Delta\zeta_i}^*$ by numerical matrix inversion. From the result we
evaluate the mean $\langle\Delta\zeta^*\rangle=|\sum_{i=1}^{NM}
{\Delta\zeta_i}^*|/NM$ and the 
root mean
square $\langle\Delta {\zeta^*}^2\rangle^{1/2}=(\sum_{i=1}^{NM}
|{\Delta \zeta_i}^*|^2)^{1/2}/NM$,
deflection amplitude (see discussion below).
The results are shown in
Fig.\ref{fig:d6p15H} with the solid and dashed line, respectively.
For the moment note that both curves are nearly identical up to
frequencies well above the lowest resonance. The theoretically
obtained resonance frequency $f_c=159$kHz is in excellent
agreement with the experimental data, showing
that collective effects are crucial for the dynamics of the
system. Moreover, the good agreement shows that monopole
interaction is effective beyond its obvious range of applicability
where $a\gg R$.

\begin{figure}
\includegraphics[bb=86 96 326 270,scale=0.5]{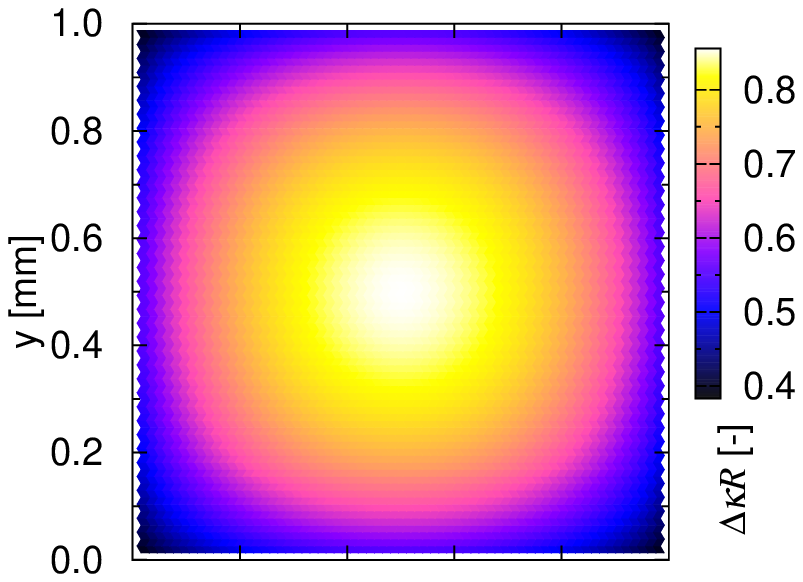}\rput(-44mm,28mm){a)}
\includegraphics[bb=112 96 326 270,scale=0.5]{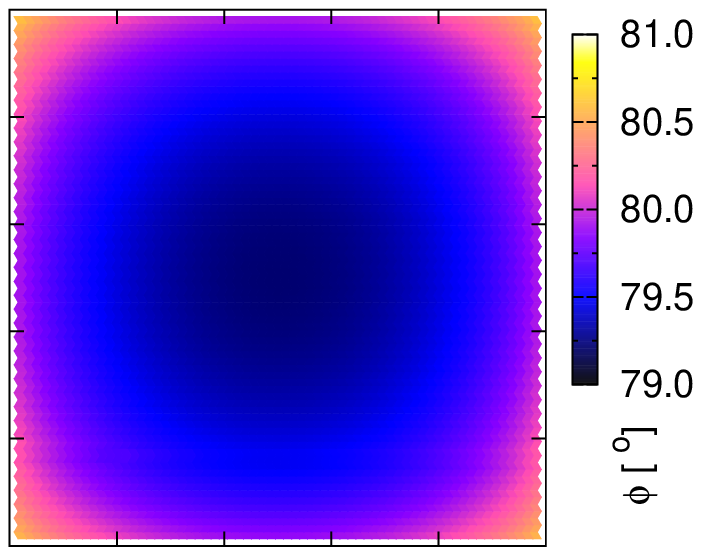}
\includegraphics[bb=86 66 326 270,scale=0.5]{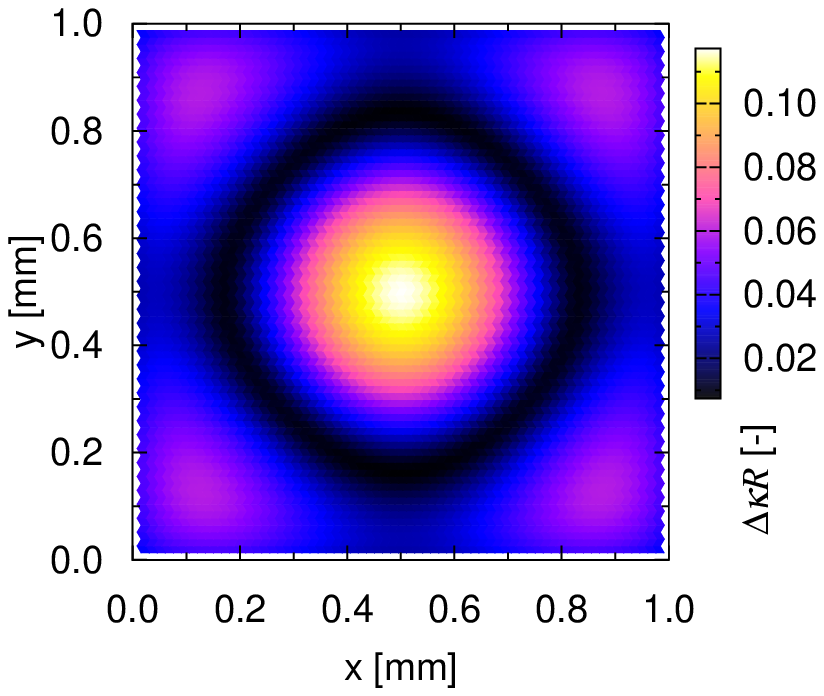}
\rput(-44mm,33mm){b)}
\includegraphics[bb=112 66 326 270,scale=0.5]{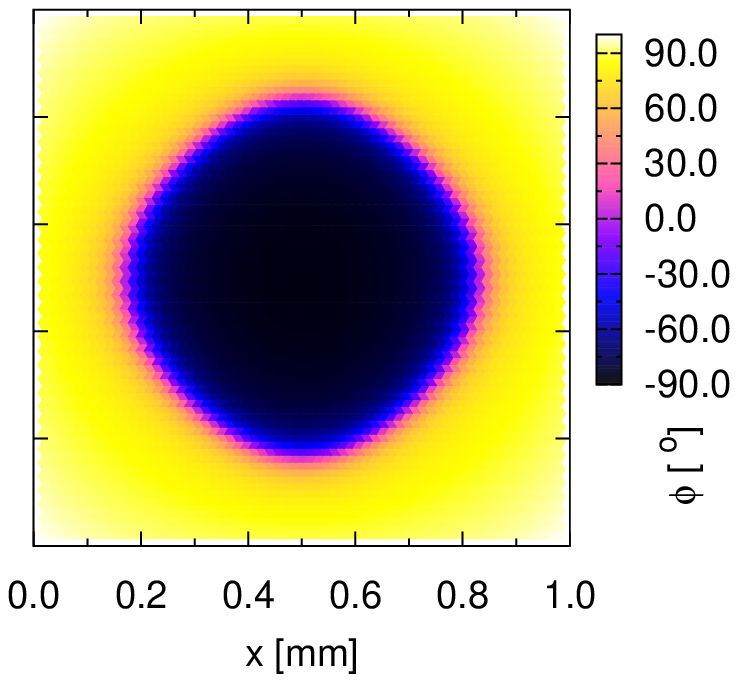}
\caption{Amplitude and phase of an array of
micromenisci with hexagonal pattern, $a=15\mu$m and $R=3\mu$m
a) at the
 fundamental collective mode (159kHz), and b)
at the second collective mode
(290kHz).
Note that the color scale for the phase is narrower in a) as compared
to b).}
\label{fig:d6p15H_phase}
\end{figure}

As can be seen in the inset of Fig.\ref{fig:d6p15H}, the theory
overestimates
the amplitude of the resonance, presumably due to the neglectance of
the bulk dissipation that arises in the collective flow
field. Second, and
more interestingly, the theoretical data for the root mean square
displays a second resonance at 290kHz which is absent in the
experimental data.
To understand the latter effect we plot the
calculated amplitude and phase for
the two lowest resonances as a function of the meniscus
position within the array in Fig.\ref{fig:d6p15H_phase}. For the
lowest resonance all menisci oscillate essentially in phase, whereas
at the second resonance, the menisci
in the center of the array and the
ones along the edge oscillate 180$^\circ$ out of phase and the
amplitude displays a node at the ring shaped boundary between the two
regions. To evaluate the diffraction intensity for such arrays
of non-identical scatterers, we note that the variation of the
meniscus deflection gives rise to phase differences between the waves
emitted from different unit cells that are much smaller than $2\pi$ --
owing to the particular experimental condition assuring that the
diffracted intensity is linear in the
meniscus displacement. Extending the above Fraunhofer picture,
one shows that the diffraction
intensity is linear in the individual menisci deflections, and thus the
experiment measures the mean deflection. Thus, we have to compare the
experiment to the theoretical mean deflection, where the second
resonance is indeed nearly invisible. Note that the expression
for the mean as given above accounts for the phase since the
deflection amplitudes $\Delta\zeta_i^*$ are complex.

The reduction of the resonance frequency predicted by the theory is
confirmed for all surface patterns investigated in experiment (see
Table~\ref{tab:comparison}).  Since the coupling between the menisci
is inversely proportional to the distance, the frequency reduction is
more pronounced for smaller lattice constants. Similarly, it
is more pronounced for the hexagonal lattices than for the
square ones, owing to the larger number of nearest neighbors. At this
moment we have no clear explanation for the
slight overestimation of the resonance frequency for the
samples with smaller radius.

In conclusion, highly mobile micromenisci are present at the textures of
superhydrophobic surfaces. Their
dynamics are determined by collective modes with resonance frequencies
that
are much smaller than the resonance frequency of a single isolated
meniscus. Superhydrophobic surfaces with
particularly large contact angles or large slip length are
expected to show the lowest resonance frequencies. Optical diffraction
has proven
an accurate tool to study superhydrophobic surfaces and their
nanoscopic hydrodynamics. It remains a challenge to extend the
optical diffraction technique to the accurate study of the
microscopic shape of the menisci, presumably by measuring also
the angular dependence of the diffraction pattern and subsequently
solving the inverse diffraction problem, following upcoming ideas
in theoretical diffractive optics.

We gratefully acknowledge stimulating discussions with A.~Prosperetti
and M.~Sbragaglia and experimental support by N.~Bremond.
This work was supported through the DFG (Grant No.\ MU 1472/4).


\end{document}